\documentclass[a4paper,twoside]{article}
\usepackage{epsfig}
\usepackage{subcaption}
\usepackage{calc}
\usepackage{amssymb}
\usepackage{amstext}
\usepackage{amsmath}
\usepackage{amsthm}
\usepackage{multicol}
\usepackage{pslatex}
\usepackage{apalike}  
\usepackage{algorithm2e}
\usepackage[bottom]{footmisc} 
\usepackage{url}
\usepackage{etoolbox}
\usepackage{SCITEPRESS}  

\newcommand{\xn}{\noindent}
\newcommand{\xv}{\vspace{0.1cm}}

\newcommand{\algspec}{algorithm specification\xspace}
\newcommand{\alghspec}{algorithm-specification\xspace}
\newcommand{\alg}{algorithm\xspace}
\newcommand{\algs}{algorithms\xspace}
\newcommand{\spec}{specification\xspace}
\newcommand{\specs}{specifications\xspace}

\newbool{anonymous}	
\setbool{anonymous}{false}

\begin{document}

\title{A Specification's Realm: \\  Characterizing the Knowledge Required for \\ Executing a Given Algorithm Specification}

\ifbool{anonymous}{}
{
\author{\authorname{Assaf Marron\sup{1}
and David Harel \sup{1}}
 \affiliation{\sup{1}Department of Computer Science and Applied Mathematics, \\ Weizmann Institute of Science, Rehovot, 76100, Israel}
\email{\{assaf.marron, david.harel\}@weizmann.ac.il}
}}

\vspace{-1.0cm}


\abstract{
An \algspec in natural language or pseudocode is expected to be clear and explicit enough to enable mechanical execution.
In this position paper we contribute an initial characterization of the knowledge that an executing agent, human or machine, should possess in order to be able to carry out the instructions of a given \algspec as a stand-alone entity, independent of any system implementation.  
We argue that, for that \algspec,  such prerequisite knowledge, whether unique or shared with other specifications, can be summarized in a document of practical size.
We term this document the \emph{realm} of the \algspec. The generation of such a realm is itself a systematic analytical process, significant parts of which can be automated with the help of large language models and the reuse of existing documents. 
The \alghspec's realm would consist of specification language syntax and semantics, domain knowledge restricted to the referenced entities, inter-entity relationships, relevant underlying cause-and-effect rules, and detailed instructions and means for carrying out certain operations.
Such characterization of the realm can contribute to methodological implementation of the \algspec in diverse systems and to its formalization for mechanical verification. 
The paper also touches upon the question of assessing execution faithfulness, which is distinct from correctness: in the absence of a  reference interpretation of natural language or pseudocode specification with a given vocabulary, how can we determine if an observed agent's execution indeed complies with the input specification.
}


\onecolumn \maketitle \normalsize \setcounter{footnote}{0} \vfill


\section{Introduction} 

In philosophy of computer science,
definitions of \emph{an \alg} commonly require that instructions  be explicit and precise, enabling a human or machine agent to execute the \algspec without applying ingenuity. For \algs specified in Turing Machines, Abstract State Machines~\cite{gurevich2000seqASM},  recursors~\cite{moschovakis1998recursors}, or as the source code for a working computer program, 
such explicitness and precision are built into the definition of the formalism. The knowledge and skills of the agent are well defined, covering basically the syntax and semantics of the formalism and a respective execution environment, with memory, processing facilities, etc.

\begin{figure*}[t] 
	\centering
  \hrule
  	\includegraphics[scale=0.5]{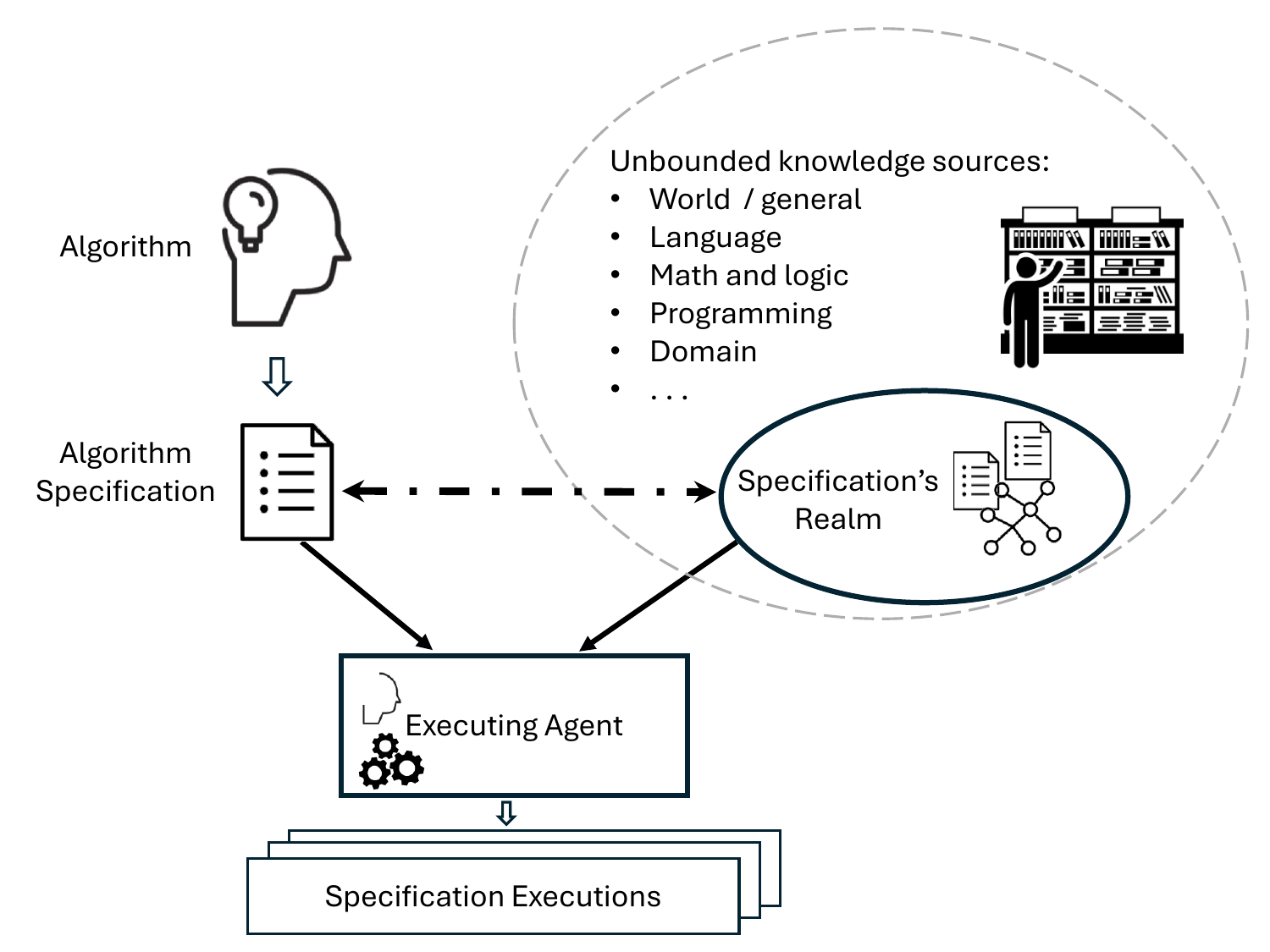}
	\caption{The concept of an algorithm specification's \textbf{realm}.
    A mental conception of \textbf{an algorithm} (top left) is concretized in \textbf{a specification document}. 
     The specification's \textbf{realm}, constructed specially for that specification as described in Sec.~\ref{sec:process}, contains the knowledge required for executing that specification.
    The realm is only a \textbf{finite subset of the unbounded information} available for the various related disciplines. 
    The executing agent --- human or machine --- receives both the \algspec and the realm constructed for it,
    and can then \textbf{execute the specification repeatedly} on diverse inputs.
    }   
  \xv 
  \xv
  \hrule
	\label{fig:Schema}
	
\end{figure*}

However, for \alg \specs in natural language (NL) 
or in pseudocode, one senses that what makes them clear to execution agents would vary significantly with the domain,  specification style, or choice of agent. Furthermore, describing what is needed for such clarity in any particular case 
remains tacit.  
Here, we contribute the following: (i) introducing the concept termed \emph{the realm of an \algspec}, which refers to the prerequisite knowledge and skills needed by an executing agent, beyond the \algspec artifact itself, to enable  execution that is ``mechanical'', i.e., not requiring ingenuity;
(ii) an initial characterization of the kinds of information such realms contain;  and (iii) a process for documenting, for any \algspec, its corresponding realm.   
Throughout we use the term \algspec to refer to an artifact embodying an \alg; this artifact and term are distinct from the terms \emph{specification} or \emph{specifications} when used in system and software engineering for documents articulating requirements or properties of the system, its behavior, or its outputs.  Fig.~\ref{fig:Schema} illustrates the relations of the algorithm, its specification, the specification's realm, and the executing agent. 

In Sec.~\ref{sec:realm}, 
we cover several categories of realm contents: language syntax and semantics of the \algspec; resolution of entity references; relevant inter-entity relationships and cause-and-effects rules; and tools for carrying out certain specified actions. As the realm concept evolves, this   categorization may be refined or extended.

Most importantly, for a given \algspec, its realm is finite and bounded: it contains only a small subset of 
the unbounded information about natural language, mathematics, programming, and diverse engineering and real-world domains that a single hypothetical general-purpose
\alg-\spec-execution agent, human or machine, would have to possess.

Documenting an \alghspec's finite realm can help guide  its implementation  in diverse systems and simplify 
the  analysis of its complexity, behavior patterns, special cases, etc. The realm can also assist in formalizing the \spec and its assumptions and goals toward formal verification of correctness.  
Finally, unions of a finite number of \algspec  realms in a given domain, might serve in constructing domain-specific agents for executing not-previously-seen {\algspec}s. 

Throughout this paper, the \alg at hand is assumed to be documented in a physical artifact $S$ ---  the \alg's \spec~--- written in NL or pseudocode; we may refer to ``the \alg'', expecting the distinction between the concrete \specs and the abstract concept of a named algorithm, like Quicksort or Binary Search, to be understood from the context.   

\begin{figure*}[t] 
  	\centering
    \includegraphics[scale=0.62]{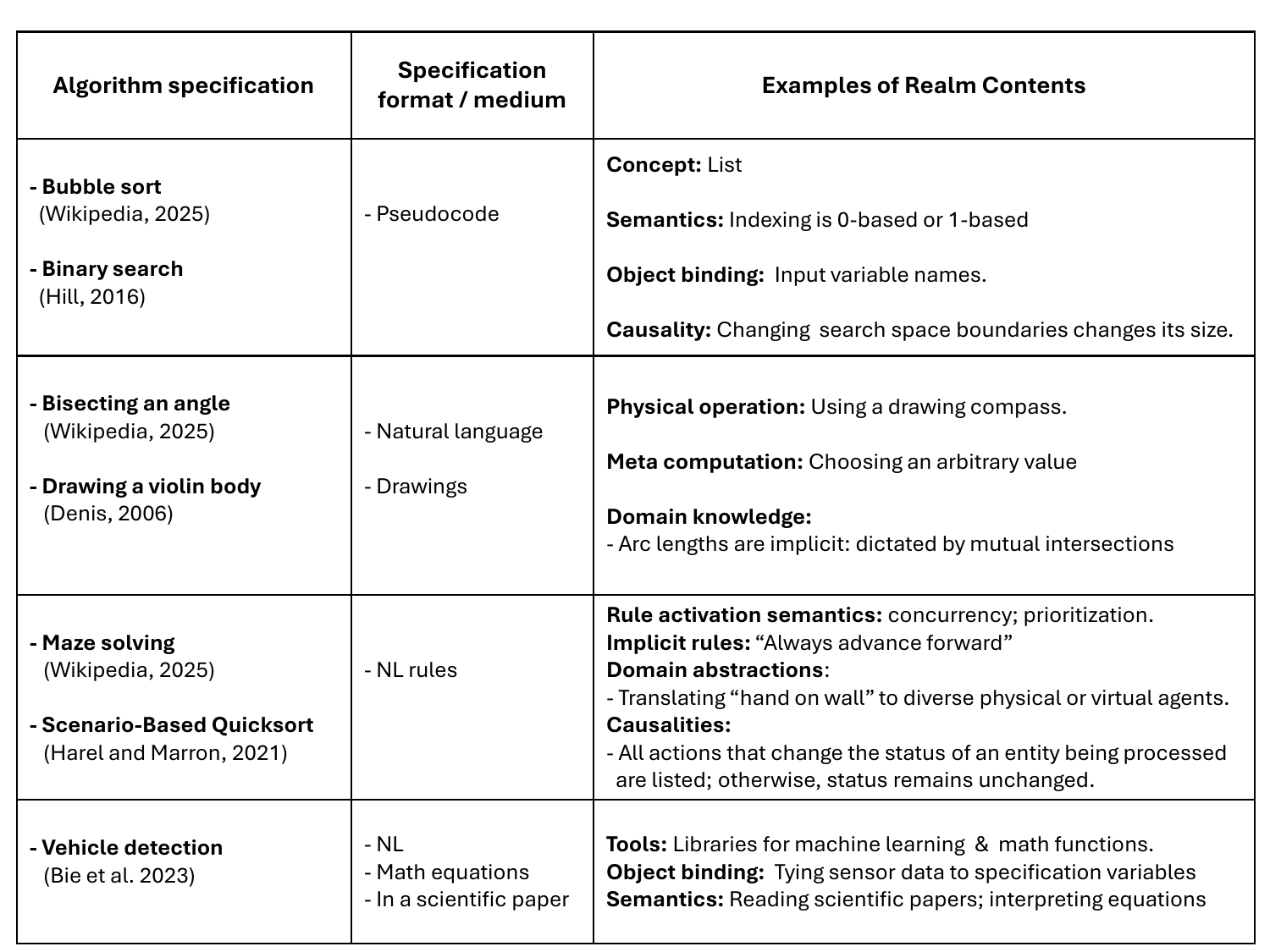}
	\caption{Examples of {\algspec}s and their realms. For each \algspec, identified with its source in the first column, the second column describes the style and format of the \spec.   The third column contains  informal statements of a few examples of additional facts and skills that are not part of the \algspec, and must be established in the \spec's realm in order to enable an agent to execute it. } 

    \vspace{0.2cm}
  \hrule
	\label{fig:AlgorithmsAndRealmsExamples}
\end{figure*}

To set the stage for our discussion,  Figure~\ref{fig:AlgorithmsAndRealmsExamples} lists several diverse \alg \specs, and a few examples of the kinds of information their realms would contain. 
Due to space considerations and to retain the flow, we rely on readers' prior familiarity with most such types of \specs and omit images of concrete \spec examples. Beyond classic compact sequential pseudocode \specs and an entire written article describing an \alg for solving a mathematical or technical problem, like~\cite{bie2023vehicleDetection}, we highlight two additional styles: 
One is rule-based or scenario-based \spec. The executing agent has to apply a collection of self-standing rules subject to pre-determined semantics. For example, in the right hand rule for maze solving, in order to exit the maze, a human agent must, constantly abide by two rules (a) keep a hand on the right wall, and (b) proceed forward. In the approach described 
in
\ifbool{anonymous}
{[X1]}
{~\cite{harelMarron2018SBAjuraj,harelMarronYerushalmi2021SBAieee}}
an intricate algorithm, like Quicksort, which is often described in a procedural step-by-step manner, possibly with recursive calls, is specified as a collection of self-standing scenarios, each of the form \emph{``Always,  when \texttt{<condition>} holds, do \texttt{<action>}''}. The choice of how  the various scenarios are orchestrated is a foundational element of the general approach, and may vary across diverse scenario-based formalisms. 

In yet another style, \specs detail artisans' instructions for activities in their trade, say, the making of violins;  such instructions may have been passed through generations in various forms, and may rely on extensive tacit domain-specific knowledge and practices and on professional jargon~\cite{denis2006traiteLutherie}. 

In discussing what an executing agent may or may not know in advance, we rely on the assumption that an executing agent is able to process new well-specified information, and learn new well-defined skills, and apply these in the mechanical execution of an \algspec. 

This assumption is needed, as even the most basic skills may be domain specific, or even algorithm specific. Considering the examples in Figure~\ref{fig:AlgorithmsAndRealmsExamples}, the maze algorithm does not require arithmetic, the drawing of a violin curve does not require understanding arrays or computer memory, and binary search does not require understanding the geometry of a two-dimensional plane with paths and obstacles. We also defer answering whether certain skills or knowledge are common to all agents, or must exist in all algorithm realms, or represent another category of knowledge altogether. Note that the assumed meta-ability of an agent to acquire foundational skills ad-hoc, or interpret additional information, appears quite advanced. It may be accomplished by hand-crafted programming of the agent by humans or by incorporating  artificial intelligence and machine learning tools like large language models in the agent. A child or a university student may or may not possess this meta-ability naturally, and may or may not be able to acquire it. 

Indeed, even the ability to follow instructions is not a trivial one. Sequential step-by-step processing may appear intuitive, but following nested iterations (loops) or go-to instructions is harder. So is concurrent application of rules, with  the agent having to understand issues like order of rule evaluation, rule side effects, action selection,  concurrent execution of physical operations, and the semantics of continuous versus discrete and instantaneous  operations. 

Hence, while the realm concept aims at enabling mechanical execution, we refrain from asserting whether an execution relies on ingenuity or not. 

The paper is structured as follows. In Sec.~\ref{sec:defs} we briefly review the known requirement that an algorithm specification be executable without applying ingenuity. 
In Sec.~\ref{sec:realm} we introduce the kinds of elements that comprise the realm. 
In Sec.~\ref{sec:process} we outline a practical constructive process whose input is an algorithm specification and its output is a document containing the realm for that algorithm.  
In Sec.~\ref{sec:relwork} we discuss related work.
In Sec.~\ref{sec:discussion} we discuss the significance of recognizing the concept of an algorithm's realm and several other implications of the current work.  

\section{The Demand for Clarity}\label{sec:defs} 

The definition of the general concept of an algorithm draws much discussion, see, e.g., 
\ifbool{anonymous}
{[X2],}
{~\cite{harel1987algorithmics},}
~\cite{hill2016alg},\cite{papayannopoulos2023algorithms},\cite[Ch.6]{primiero2020foundationsComputing}. 
In this paper we sidestep most of the subtleties associated with the original definitions, their respective contexts and subsequent discussions. Instead, we focus on characterizing the knowledge required of the agent executing an \algspec. Still, as a starting point for the current discussion, we present below selected definitions as cited or paraphrased in~\cite{hill2016alg}. 
In each paragraph, we  highlight with italics words and phrases that imply the requirement that the instructions be clear to the agent, sufficiently to enable execution. 

\xv 
\xn ``\textbf{Markov and Nagorny (1988)}: An algorithm is a prescription \emph{uniquely determining} the course of certain constructive processes.'' 

\xv
\xn ``\textbf{Kleene (1967):} If (after the procedure has been described) we select any question of the class,
the procedure will then \emph{tell us how} to perform successive steps, after a finite
number of which we will have the answer to the questions we selected. In performing
the steps, \emph{we have only to follow the instructions mechanically, like
robots; no insight or ingenuity or invention is required of us}. After any step, if
we don’t have the answer yet, the instructions together with the existing situation
will tell us what to do next. The instructions will enable us to recognize
when the steps come to an end, and to read off from the resulting situation the
answer to the question, “yes” or “no”. In particular, [......], the description of the procedure,
by a list of rules or instructions, must be finite.''

\xv 
\xn ``\textbf{Minsky (1967):} 
An \emph{effective} procedure is a set of \emph{rules which tell us}, from moment to moment, how to behave.''

\xv 
\xn ``\textbf{Knuth (1997):} `An algorithm is a finite set of rules that gives a sequence of operations for solving a specific type of problem,' with five essential features: finiteness, \emph{definiteness}, input, output, effectiveness.''

\xv 
\xn ``\textbf{Rapaport (2012)} (synthesized definition): An algorithm (for executor E to accomplish goal G) is:
A procedure (or method)—i.e., a finite set (or sequence) of statements (or rules, or instructions)—such that each statement is: 
\vspace{-0.2cm}
\begin{itemize}
\item[-] Composed of a finite number of symbols (or marks) from a finite alphabet
\item[-] {And \emph{unambiguous} for E—i.e.,}
\begin{itemize}
\item[-]{E \emph{knows} how to do it}
\item[-]{ E \emph{can} do it}
\item[-]{it can be done in a finite amount of time}
\item[-]{and, after doing it, E knows what to do next—}
\end{itemize}
\item[-]{And the procedure takes a finite amount of time, i.e., halts,}
\item[-]{And it ends with G accomplished.''}
\end{itemize}

\xv  
\xn ``\textbf{Hill (2016): Definition 1.} An algorithm is a finite, abstract, \emph{effective}, compound control structure,
\emph{imperatively} given.''

\xv  
\xn ``\textbf{Hill (2016): Definition 2.} An algorithm is a finite, abstract, \emph{effective}, compound control structure,
\emph{imperatively} given, accomplishing a given purpose under given provisions.''

\xv 
\xv 
\xn In these definitions, the demand for clarity of an \alghspec's instructions is manifest.
However, 
the characterization of what makes a \spec clear, or what an executing agent must know in advance in order to be able to follow certain instructions, remains tacit. 

\section{Contents of a Specification's Realm } \label{sec:realm}

We assume that once an executing agent $G$ is presented with an \algspec $S$, it can process its realm $R$ acquiring the additional information.  
Then, in executing $S$, $G$ can associate the various artifacts of $S$ and $R$, like objects and actions, with its internal execution templates and processing capabilities. 
In the coming subsections, we offer an initial classification of the elements in  {\algspec}s realms and delineate the realm boundaries. 

\subsection{Language syntax and semantics}
The realm contains explanations of all the language constructs that appear in the \algspec. 
They may be general and common in some language domains, or unique to the specification at hand. These include the syntax of individual statements, flow or step-by-step instructions, looping and iteration logic (including termination and break semantics), code blocks notation (begin/end, curly brackets/braces, indentation, etc.), composition of stand-alone rules, reliance on details that are explained elsewhere (as in function calls, usage of structures like \emph{"where \texttt{<this>} is \texttt{<that>}"}, pointers to  explanatory footnotes, etc.), and more.  

The realm may also contain a preamble explaining the syntax and semantics of other realm contents.

We defer to future assessment whether interpreting an algorithm provided in a multi-page scientific paper is a basic agent skill or should be part of a realm. This includes relying on a host of software libraries and mathematical notations, and the ability to assemble a sequential process from a  collection of sub-processes referencing each other in diverse ways.

\subsection{Foundational Model Entities}

An \algspec contains or implies a model of the context in which it operates; where applicable, this context may be thought of as a system and its environment. Such a model includes objects (as understood in object-oriented modeling), properties, property values, object methods, conditions, states, and events. The model also includes actions that are not directly associated with objects, like meta-operations regarding the execution environment, meta-computation like choosing a random value, entity creation and deletion, establishing relations among existing entities, etc. 

A realm $R$  of an \algspec $S$ should:
(i) identify which terms in $S$ serve as 
objects, properties, methods, events, other actions, etc.; 
(ii) unify synonymous references to the same entity; 
(iii) disambiguate cases where the same terminology refers to distinct entities; and (iv) elucidate any implicit or absent entity references making them explicit.

\subsection{Entity Relationships} 

The realm should use domain knowledge to fill in entity relationships, like \emph{contains}, \emph{part-of}, or \emph{kind-of}, that are relevant to execution. 
For example: that a list $A$ to be sorted \emph{contains} the entities which are being compared and reordered; which arc in a violin-body circumference is \emph{connected to} which other two arcs; or, which set of traffic sensors, like for location and velocity is associated with which monitored vehicle.

\subsection{Domain Causality Rules} 

A deeper layer of information in the \alghspec realm is articulation of cause-and-effect rules that hold in the execution environment. Here are a few examples:  for the binary search algorithm in ~\cite{hill2016alg} the realm should add the implicit fact that the variable $N$ containing the search space length is updated automatically whenever the variables \texttt{Lower} or \texttt{Upper} marking the search space boundaries are updated; when carrying out bubble sort in a computer memory, storing a value in a memory cell erases irretrievably its previous contents, hence swap operations must be done with temporary holding spaces; in~\cite{bie2023vehicleDetection}, the 
the \algspec for algorithm YOLOv5n-L is a list of step-by-step modifications to the published algorithm YOLOv5n. Once these steps are completed, an agent executing  YOLOv5n-L must have access to the behavior of YOLOv5n. 

When the \algspec is part of a larger system engineering model, some causality rules may be derived from descriptions of dynamic system behavior; see Sec.~\ref{sec:relwork} for more details.

\subsection{Instruction explanations and tools}
The \algspec may include instructions that appear explicit and clear to their author, but that the agent at hand cannot carry out. The realm will thus elucidate opaque instructions, 
provide necessary instrumentation, or explain how to obtain information that is required but not provided. Examples: 
\begin{enumerate}
    \item the realm may explain to a smart and dexterous robot agent the use of a drawing compass: placement of the sharp point, measuring the radius, etc. 
    \item{for a maze-solving software agent, the realm may include a software function for converting a high-resolution maze map image into a smaller matrix capturing the abstracted, schematic maze structure; it may also explain that the ``hand-on-wall'' is a metaphor for a minimal numerical distance.}
    \item{the realm may show a human or machine agent, how, for the first time, to make a choice, as in the instruction \emph{"pick an arbitrary radius and draw a circle around point C."} (as appears in some algorithms for bisecting an angle). In such a case, the realm will also introduce value range constraints.} 
    \item{The realm may explain to an agent drawing a violin body contour that the lengths of the bouts (=arcs) are not specified; instead they are implied by the intersection points of adjacent circles.} 
\end{enumerate}

\subsection{Realm scope and boundaries}
The realm is provided in order to enable mechanical execution. 
Including additional information about the domain or the problem, the algorithm's specific solution, and perhaps about alternative solutions, is not needed and should preferably be excluded. 
For example, if the \algspec is for carrying out binary search in a list, the realm should not include a function or library that replaces the \alghspec's steps with one or more alternative function calls. The \algspec defines a particular choice of steps, in a particular order or compositional semantics. Furthermore, the specification may be focused on contrasting this choice with other \algspec{s} for solving the same problem, rather than on explaining for the first time how the problem should be solved. This choice should remain explicit, and the realm should not  hide or obscure it. 

\section{Constructing a Specification's Realm} \label{sec:process}

Below we provide an outline of one possible manual and computer assisted structured process for creating a document containing a realm $R$ for a given \algspec $S$. 
Our goal here is to emphasize the existence of such processes, and the finiteness and  manageability of their output artifacts. Clearly, with present and future large-language-model tools, many of these steps can be readily automated. 
The description of some of these steps can be aligned with steps in classic object modeling and requirements engineering guidelines, as in~\cite{booch2005UMLUserGuide,deeptimahanti2009automatedUMLNounVerbRUP}, although the overall process is very different, as discussed in Sec.~\ref{sec:relwork}.

\begin{enumerate}

\item{If $S$ is in pseudocode aligned with a particular programming language - add to $R$ a language reference manual describing the syntax and semantics of that language. Add paragraphs as needed to highlight differences. You can suffice with those aspects of the language and pseudocode that are used in $S$.}
\item{Parse $S$; extract nouns, verbs and adjectives, creating a glossary. If $S$ is in NL, you may use a large language model (LLM) for this.}
\item{After each step below, examine $R$'s vocabulary subject to domain knowledge: equate synonyms, add qualifiers to disambiguate similar distinct references, and concretize implicit references.}
\item{Using domain knowledge, possibly with the help of an LLM, transform the above glossary into an initial model $M$ class diagram of objects and their properties and methods, as well as a list of events.}
\item{Associate allowed property values with the various properties.}
\item{Identify entity references that depend on relationships between entities. Use domain knowledge to complete $M$ with these relevant relationships.}
\item{Determine which invocation of a method call, property change or other action  triggers each event that appears in the emerging model $M$ . Make sure that such triggers are indeed specified directly in $S$ or are specified as a result of causality rules in $R$. When no such cause-and-effect rule is found, use domain knowledge to fill that gap. For each new causality rule, add its triggering conditions and events to $M$ and repeat the above steps to fill any gaps introduced by the new entities.}
\item{Examine the executing agent's capabilities. For each method call, property change, and, most importantly, action of some other kind, that appears in $M$, check  if the agent is able to carry it out directly. If not, provide the necessary information, like adding a library and an appropriate API, or a procedure  for accomplishing the same task, consisting of substeps that the agent can carry out.}  
\item{Finally, examine $S$ and $R$, and for each entity, ask  whether the agent knows what it is and/or how to handle it. When the answer is no, add a layer of entities, causality rules, and bindings to entities in $S$ and $R$ sufficient for executing $S$. The decision of when to stop a realm construction process can be understood in the context of vertical and horizontal delineation, termed in
\ifbool{anonymous}
{[X3]}{
~\cite{harel1987statecharts}}
simply as depth and modularity. 
These are, respectively, the limits forced on the top-down depth in refining  hierarchical abstractions, and the scope or contents of a system or a module relative to everything else. In classic modeling, these traits  are dictated by an outside-in view of a mix of the known requirements and implementation resources and constraints. By contrast, realm construction induces an inside-out view: the vertical depth and the horizontal scope of the specification's realm are dictated by the content of $S$, and the capabilities of the chosen execution agent $G$. 
} 
\end{enumerate}

Note that LLMs may be used in the construction of $R$  but are not used in the execution of $S$ given $R$.


\section{Related Work}\label{sec:relwork}

The concept of an \alghspec's realm differs from the related concept of Problem Frames~\cite{jackson2005problemFrames} as follows. 
Realms embody an inside-out view: starting with the specification, missing elements are filled in from multiple dimensions: the entities involved and their relationships as needed for execution,  behavior of the tacit, possibly synthetic  model of the environment to which the \algspec  is targeted,  and the syntax and semantics of the specification artifact. The parts of the realm that relate to real-world implementation may then be further mapped to the entities of the intended implementation. 
By contrast, in problem frames the starting point is the real world, and the physical application of the system being developed. Constructing a well-defined model for these, with a consistent terminology can help create useful specifications and detailed models, and eventually, a final working system. 

Similarly, as stated in Sec.~\ref{sec:process}, when constructing a realm for enabling the execution of a specification, lower level activities may be reminiscent of activities in modeling system structure and behavior from initial specifications~\cite{booch2005UMLUserGuide,deeptimahanti2009automatedUMLNounVerbRUP}. However, the bigger pictures of  the inputs and outputs of the modeling process, and subsequent use of the constructed model are different. In system and software engineering, modeling is part of the design and development of a system; the model permeates all aspects of the system and of its development process, where an \alghspec's realm is an essential synthetic artifact, which may be detached from any particular system. A realm is governed by an existing \algspec for the purpose of executing it.  

One may argue that an agent equipped with a powerful LLM may be able to execute many \alg \specs, or generate code that does so, without depending on any realms, as it may have all the necessary world and domain knowledge (contents of the dashed circle in Figure~\ref{fig:Schema}). We believe that identifying what knowledge the agent actually uses and depends on is critical for analyzing the \algspec and individual executions (See Sec.~\ref{sec:discussion} below).

\section{Discussion} \label{sec:discussion}

\xv 
\xn \textbf{Significance.} First, introducing the concept of the realm of an \algspec, and the various types of knowledge and skills its execution assumes or requires,  elucidates the general concepts of an \alg, \spec, and execution.  

Second, having a realm for a given \spec assists in developing implementations in diverse systems. Moreover, whether or not LLMs are used in realm construction, articulating the tacit information may help in code generation, formalization for verification, and other \alg analysis activities.  
For example, in preparing for formal verification, the elements defined in the \algspec and its realm may provide an initial inventory of the states and transitions that constitute the state-transition system underlying the model to be verified~\cite[Ch.2]{baierKatoen2008ModelChecking}. Naturally, the model to be verified must specify additional aspects, such as desired ("good") and undesired ("bad") states and terminal or stopping states, and it may  reflect further refinement or abstraction of the raw elements in the \algspec and its realm. 

Third, systematic approaches for constructing a \spec's realm, for which we have presented a preliminary outline, can be integrated into methodologies for requirements engineering, object-oriented modeling, and the modeling stage in formal methods. 

Fourth, the realm contents may contribute to interpretability of an \algspec or explainability of its actual behavior. Given that {\algspec}s may themselves be computer generated (e.g., by LLMs), the insights added by the realm can become indispensable.  

Finally, it will be interesting to investigate the applicability of the concept of realm as a complement of a narrow specification in constructing test environments for individual software modules and components in system development. 

\xv 
\xn \textbf{Faithfulness of Execution.} 
Consider an execution log or real-time observation of an arbitrary agent $G$ executing an arbitrary \algspec $S$, with or without a realm, specified in NL or pseudocode. 
An external observer --- perhaps the very author of $S$, may determine whether the steps taken by $G$ are as intended. 
Assessing faithfulness of execution is distinct from checking correctness of the \spec, which is associated with goals, pre- and post-conditions for the overall execution, and related properties.

When the \algspec is in a well-defined formalism such as a programming language or a Turing machine, faithfulness is implicit and guaranteed by the formal semantics. 
For an \algspec in NL or pseudocode, one possible approach could call for associating every imperative instruction with pre- and post-conditions defining its intended effects. Nevertheless, while the goal of constructing the realm is to enable faithful execution, in this paper we only highlight the issue, and defer to future research the question of how to define and how to measure faithfulness.

\xv 
\xn \textbf{Repetitive Execution.} 
Definitions of the concept of an algorithm imply that an agent executing an \algspec can carry out the execution again and again --- on the same or diverse inputs. We observe that, for an \algspec $S$, the construction of its realm $R$ is required only once.

\xv 
\xn \textbf{Realm Depth.}
The process of constructing the realm $R$ of an \algspec $S$ outlined in Sec.~\ref{sec:process} includes an iterative questioning approach as in ``what is~\texttt{<this>}'',  which may be applied to almost any element of $S$ and  $R$, and thus continue indefinitely.
Furthermore, the basic skills of the executing agent may contain patterns that the author of $S$ does not care about, or is willing to accept common defaults for, and thus wishes to stop the flow of questions. For example, realm construction may yield the question: ``to what precision should the agent compute the real number $V$?'', or ``what rounding protocol should be used when complying with the desired precision?''. 

We propose several approaches for addressing this dilemma: 
(1) Continue the process until the binding of the elements of $S$ and $R$ to the capabilities of the agent can be complete, possibly using default values. Some such defaults may not be known and are left to the agent at execution time. Such choices can then be evaluated in the context of execution faithfulness mentioned above. 
(2) Reexamine the choice of agent! A more capable agent or a more na\"ive one may be more suitable for the execution of $S$. 
(3) Revise $S$. The challenging questions when constructing the realm may highlight the fact that the \algspec is indeed deficient and requires work. 

\xv 
\xn \textbf{Realm practicality.} 
One may wonder whether a well-constructed realm, while finite, will be too large to be useful. Here the answer is simple: while human engineers may study the realm and learn from it, it is a resource for \spec execution agents, and LLM-based code generation tools. For these, the size of the realm is less of an issue. Also, in many domains realms may be shared across multiple \specs documents.  Such reuse then affects not only each realms' sizes, but also the effort to construct them.


\section{Conclusion and Future Research}\label{sec:conclusion}

In this position paper,  we have introduced the concept of the 
realm  of an \algspec. 
The realm elucidates and expands the specification with the domain and general information and tools needed for mechanical execution. 

Once constructed, the realm can be used in repeated execution, and is a useful resource for implementations and for formal verification. 
Future research directions include producing detailed, comprehensive realms for a variety of algorithms to serve as a reference for the concept;  refining the categorization of realm contents; developing tools for automated realm construction; and establishing a structured process for using an \alghspec's realm for constructing a formally verifiable model. 
The realm concept emerged from the study of the fundamental concept of an algorithm and its specification; it will be interesting to explore the applicability of such realms  in diverse areas in system and software engineering.

\ifbool{anonymous}{}
{
\subsubsection*{Acknowledgments} 
We thank Kenneth Beckmann for insights and references about algorithms in the art of violin making. We thank Orley K. Marron for valuable discussions and suggestions. 
This research was funded in part by research grants to DH from Louis J. Lavigne and Nancy Rothman, the Carter Chapman Shreve Family Foundation, Dr. and Mrs. Donald Rivin, and the Estate of Smigel Trust.}

\bibliographystyle{apalike} 

\end{document}